\documentstyle[prl,tighten,aps,epsf,graphics,multicol,amstex]{revtex}

\newcommand{\id}{{\sf 1 \hspace{-0.3ex} \rule{0.1ex}{1.52ex} \rule[-.02ex]{0.3ex}{0.1ex} }}
\newtheorem{theorem}{Theorem}
\newtheorem{definition}[theorem]{Definition}
\newtheorem{lemma}[theorem]{Lemma}

\newtheorem{corollary}[theorem]{Corollary}

\begin{document}

\draft

\title{Operator monotones, the reduction criterion and the relative entropy.}
\author{M.B. Plenio, S. Virmani, P. Papadopoulos}
\address{Optics Section, Blackett Laboratory, Imperial College, London SW7 2BZ, United Kingdom}

\date{\today}
\maketitle

\begin{abstract}
We introduce the theory of operator monotone functions and employ
it to derive a new inequality relating the quantum relative
entropy and the quantum conditional entropy. We present
applications of this new inequality and in particular we prove a
new lower bound on the relative entropy of entanglement and other
properties of entanglement measures.
\end{abstract}


\begin{multicols}{2}
\narrowtext Recently the entanglement of finite systems has
received considerable attention \cite{finite,Nielsen} when it was
realized that the theory of majorization \cite{Marshall,Bhatia}
provides a simple mathematical framework in which the theory can
be formulated \cite{Nielsen}. In general the well developed theory
of matrix analysis provides many techniques and ideas that may be
useful for the study of entanglement. However, the restriction to
finite entanglement, while justified from an experimental point of
view, places an additional constraint on the system which may
cloud some of the truly fundamental aspects of entanglement.
Therefore the study of the asymptotic limit, i.e. a situation in
which large numbers of entangled pairs can be manipulated
simultaneously, is of interest from a fundamental point of view. A
substantial body of work has been developed in recent years,
beginning with the case of pure entangled states \cite{Bennett96a}
and an extensive study of different ways to quantify the amount of
entanglement in mixed bipartite states. Some interesting examples
are the entanglement of formation \cite{Bennett96b,Wootters1}, the
entanglement of distillation, and the relative entropy of
entanglement \cite{Vedral1,Vedral2,Vedral3,Rains}. With the
notable exception of the entanglement of formation
\cite{Wootters1}, these entanglement measures are very difficult
to compute analytically even in the qubit case. Therefore it is of
great interest to obtain upper and lower bounds for them. To
further our understanding of entanglement and our ability to
manipulate it locally, it is of interest to try to establish
connections with other ideas such as distinguishability
\cite{Donald1,Vedral2,Vedral3}, and thermodynamical considerations
\cite{Pleniohol,thermo}. In these contexts one mathematical
function emerges as a central quantity, namely the relative
entropy which is defined as
\begin{equation}\label{Eq1}
  S(\sigma || \rho) = tr\{ \sigma \log\sigma - \sigma\log\rho \} \;\; .
\end{equation}
It has a number of remarkable properties
\cite{Vedral1,Donald1,Schumi,Wehrl} and is closely related to the
problem of the quantification of entanglement
\cite{Vedral1,Vedral2,Vedral3,Rains}, the distinguishability of
quantum states \cite{Vedral2,Vedral3} and to thermodynamical ideas
(see for example \cite{Donald2}). Any new inequality relating the
relative entropy to other entropic quantities is therefore
expected to lead to potentially important new insights into any of
these topics and is potentially an important contribution.

In general one would attempt to formulate inequalities that are
valid for any density operator. For the study of entanglement,
however, a particular special type of inequality would be of great
interest. These are inequalities that are only valid when at least
some of the density operators that are involved are
non-distillable or separable, but may be violated for distillable
states. These inequalities naturally lead to sufficient criteria
for the distillability of states and they are, as we will
demonstrate here, very useful for example in the study of
entanglement measures.

In this paper we combine the ideas of positive maps
\cite{Horodecki1,Terhal} with the concept of operator monotones
which has been developed originally in matrix analysis
\cite{Bhatia} to derive such a new inequality relating the
relative entropy and the entropy. We present some lemmas and
corollaries to this inequality to demonstrate its usefulness. In
particular we derive a new lower bound on the relative entropy of
entanglement and a much simplified proof that for pure states the
relative entropy of entanglement coincides with the entropy of
entanglement.

Let us briefly introduce the idea of operator monotone function as this is a
concept which is not very familiar to quantum information theory. Much more
material can be found for example in \cite{Bhatia}. First we begin with
\begin{definition} \label{Def1}
Given two operators $A$ and $B$, we say that $A\ge B$
if the operator $A-B$ is a non-negative operator, i.e.
$A\ge B$ if for all $|\psi\rangle$ we have $\langle \psi|A-B|\psi\rangle\ge 0$.
\end{definition}
This definition allows us to compare operators and in particular
we are now able to introduce the idea of operator monotones. Given
a real valued function $f: \mathbb{R}\rightarrow \mathbb{R}$ we
canonically extend it to a function on Hermitean operators
\cite{Bhatia}. Then we make the following
\begin{definition} \label{Def2}
A function $f$ is called {\bf operator monotone}, if for
all pairs of Hermitean operators satisfying $A\ge B$ we have $f(A)\ge f(B)$.
\end{definition}
It should be noted that ordinary monotonicity of a function and operator monotonicity
are two entirely different concepts. An example is the function $f(x)=x^2$
on the interval $[0,\infty]$, which is not an operator monotone function although it
is clearly a monotone function in the ordinary sense \cite{Bhatia}! In physics, and
in particular in thermodynamics and the theory of entanglement the entropy and therefore
the logarithm plays a central role. It is therefore important to note that
\begin{lemma} \label{Lemma1}
    The function $f(x) = log(x)$ is operator monotone!
\end{lemma}
The complicated proof of this theorem can be found in \cite{Bhatia,Loewner}.
Lemma \ref{Lemma1} is one of the key ingredients in the proof of our new
inequality.

The other major input comes from the theory of positive but not completely
positive maps, whose application to quantum entanglement of mixed states was
pioneered by the Horodeckis and further developed for example in \cite{Terhal}.
Positive maps can be
used to detect the non-separability of mixed states and a number of important
positive maps have been found, amongst them the well known partial transposition
\cite{Horodecki96,Peres96}. Here we employ a different positive map which has been
introduced in \cite{Horodecki1}. This map, the reduction map $\Lambda$, is defined
as
\begin{equation}
    \Lambda(\rho) := \id\, tr \rho - \rho \;\; .
\end{equation}
The reduction map is evidently positive, but not completely positive as the
map $\id\otimes \Lambda$ is not positive, i.e. it can transform a positive operator
into a non-positive operator. The corresponding reduction criterion is then given by
\begin{equation} \label{Eq2}
    \mbox{$\rho$ is non-distillable} \Rightarrow \rho_A\otimes \id \ge \rho_{AB}
    \;\; . \label{reduction}
\end{equation}
The reduction criterion is remarkable as its violation implies distillability
of the density operator $\rho_{AB}$ while this is not known to be the case for
the partial transposition.

Now we use the two key properties of operator monotonicity of the logarithm
(Lemma \ref{Lemma1}) and the reduction criterion Eq. (\ref{Eq2}) to prove
\begin{theorem} \label{Theorem1}
For any non-distillable state $\rho_{AB}$ and for any state $\sigma_{AB}$
of a bipartite system AB we have
\begin{eqnarray}\label{EqTheorem}
    S(\sigma_A) - S(\sigma_{AB}) &\le& S(\sigma_{AB}||\rho_{AB})- S(\sigma_A||\rho_{A}) \; , \\
    S(\sigma_B) - S(\sigma_{AB}) &\le& S(\sigma_{AB}||\rho_{AB})- S(\sigma_B||\rho_{B}) \; .
\end{eqnarray}
\end{theorem}
It should be noted that the left hand side of the inequality is the negative
conditional entropy which is negative for all separable states $\sigma_{AB}$, while
it can take positive values for entangled states (an example is the singlet state).

Before we discuss the implications of this theorem further let us present its proof.\\
\noindent
{\bf Proof:} Given a non-distillable state $\rho_{AB}$, the reduction criterion
and the operator monotonicity of the logarithm imply that \cite{Cerf}
\begin{displaymath}
    \log(\rho_A\otimes \id_B) \ge \log(\rho_{AB}) \;\; .
\end{displaymath}
This statement is equivalent to
\begin{eqnarray*}
    &&\forall \sigma_{AB}: \; tr\{\sigma_{AB} \log(\rho_A) \otimes \id_B\} \ge
    tr\{\sigma_{AB} \log \rho_{AB}\} \\[0.1cm]
    &\Leftrightarrow& \forall \sigma_{AB}:  - tr\{\sigma_{AB} \log(\rho_A) \otimes \id_B\} \le -
    tr\{\sigma_{AB} \log\rho_{AB}\}\\[0.1cm]
    &\Leftrightarrow& \forall \sigma_{AB}:
    - tr\{\sigma_A \log\rho_A\} \le - tr\{\sigma_{AB} \log\rho_{AB}\} \; .
\end{eqnarray*}
To draw the connection to the relative entropy we use Definition \ref{Eq1}
to find the equivalent statement
\begin{eqnarray*}
    &&\forall \sigma_{AB}:
    - S(\sigma_{AB}) +  S(\sigma_A) - S(\sigma_A) - tr\{\sigma_A \log\rho_A\}
    \le\\ && \;\;\; - S(\sigma_{AB}) - tr\{\sigma_{AB} \log\rho_{AB}\}\\[0.1cm]
    &\Leftrightarrow& \forall \sigma_{AB}: - S(\sigma_{AB}) +  S(\sigma_A) +
    S(\sigma_A||\rho_{A}) \le S(\sigma_{AB}||\rho_{AB}) \\[0.1cm]
    &\Leftrightarrow& \forall \sigma_{AB}: S(\sigma_A) - S(\sigma_{AB}) \le
    S(\sigma_{AB}||\rho_{AB})- S(\sigma_A||\rho_{A}) \;\; .
\end{eqnarray*}
Interchanging the roles of $A$ and $B$ we find the second inequality $_{\Box}$\\

In the following we present some applications of the new inequality presented
in Theorem \ref{Theorem1}. Firstly, let us demonstrate that from Theorem
\ref{Theorem1} we can obtain a new lower bound on the relative entropy of
entanglement. We find
\begin{lemma} \label{Lemma2}
The relative entropy of entanglement is bounded from below by
the negative conditional entropy, i.e. for all $\sigma_{AB}$ we have
\begin{displaymath}
    E_{RE}(\sigma_{AB}) \ge \max\{ S(\sigma_A) -
    S(\sigma_{AB}), S(\sigma_B) - S(\sigma_{AB}) \}\;\; .
\end{displaymath}
\end{lemma}
{\bf Proof:} The relative entropy of entanglement is defined as
\begin{equation}
    E_{RE}(\sigma_{AB}) = \min_{\rho_{AB}\in \cal{D}} S(\sigma_{AB} || \rho_{AB})
\end{equation}
where $\cal{D}$ either denotes the set of separable states
\cite{Vedral1,Vedral3}, the set of states with positive partial transpose
\cite{Rains} or the set of non-distillable states \cite{Vedral4}. Lemma
\ref{Lemma2}
applies to both definitions and we only prove the strongest one, choosing
$\cal{D}$ to be the set of non-distillable states. Let us denote
by $\rho_{AB}^*$ the non-distillable state that realizes the relative entropy
of entanglement, i.e.
\begin{equation}
    E_{RE}(\sigma_{AB}) = S(\sigma_{AB} || \rho_{AB}^*) \;\; .
\end{equation}
From Theorem \ref{Theorem1} and the non-negativity of the relative entropy
we conclude that
\begin{eqnarray}
    S(\sigma_A) - S(\sigma_{AB}) &\le& S(\sigma_{AB}||\rho_{AB}^*)-
    S(\sigma_A||\rho_{A}^*) \nonumber \\
    &\le& S(\sigma_{AB}||\rho_{AB}^*) \nonumber \\
    &=& E(\sigma_{AB}) \;\; .
\end{eqnarray}
Interchanging systems $A$ and $B$ and combining the result yields the
Lemma \ref{Lemma2} $_{\Box}$\\

A direct consequence of Theorem 1 is a relationship between the relative
entropy of entanglement and the entanglement of formation.
\begin{corollary} \label{Corollary2} For any bipartite state $\sigma_{AB}$ we have
\begin{displaymath}
    E_{RE}(\sigma_{AB}) \ge E_F(\sigma_{AB}) - S(\sigma_{AB}) \; .
\end{displaymath}
\end{corollary}
{\bf Proof:} This follows immediately from Lemma \ref{Lemma2} because
$E_F(\sigma_{AB})\le S(\sigma_{A})$ $_{\Box}$

A remarkable consequence of this Theorem 1 is a very simple proof, that the
relative entropy of entanglement for pure states coincides with the entropy of
entanglement, i.e. the entropy of the reduced density operator of one
subsystem. This statement was first proven in \cite{Vedral3}, however, these
proofs are very complicated. Using Lemma \ref{Lemma2} this proof is simplified
considerably.
\begin{corollary} \label{Corollary1}
For pure states $|\psi_{AB}\rangle$ we find
\begin{equation}
    E_{RE}(|\psi_{AB}\rangle\langle\psi_{AB}|) = S( \rho_A ) \; .
\end{equation}
where $\rho_A = tr_B \{ |\psi_{AB}\rangle\langle\psi_{AB}| \}$.
\end{corollary}
{\bf Proof:} Up to local unitary operations we can write
$|\psi_{AB}\rangle= \sum_{i=1}^{n} \alpha_i |i\rangle_A |i\rangle_B$ for an
orthonormal basis $\{|i\rangle\}_{i=1,n}$. For the mixed state
$\rho_{AB}= \sum_{i=1}^{n} |\alpha_i|^2 |i i\rangle\langle i i|$ we find
$E_{RE}(|\psi_{AB}\rangle\langle\psi_{AB}|) \le S(|\psi_{AB}\rangle\langle\psi_{AB}|\, || \rho_{AB}) =
S(\rho_A)$. On the other hand Lemma 1 provides
$E_{RE}(|\psi_{AB}\rangle\langle\psi_{AB}|) \ge S(\rho_A)$ and therefore we
conclude that Corollary \ref{Corollary1} is correct $_{\Box}$

It is interesting to note that, to our knowledge, all states for which we
know the distillable entanglement under local operations and one-way
communication, it actually coincides with the negative conditional entropy and
one may conjecture that indeed the distillable entanglement under local
operations and one way communication is equal to the negative conditional
entropy. A similar conjecture has been made by Rains for maximally correlated
states \cite{Rains}.

Another small conclusion we can draw from Theorem 1 is the following
\begin{lemma} \label{Lemma4}
For states $\sigma_{AB}$ such that
    $E_{RE}(\sigma)=\max\{ S(\sigma_A) -
    S(\sigma_{AB}), S(\sigma_B) - S(\sigma_{AB})\}$
the closest state $\rho_{AB}^{*}\in\cal{D}$ must have the same reduced
density operator as $\sigma_{AB}$.
\end{lemma}
{\bf Proof:} Without restricting generality we can assume
$E_{RE}(\sigma_{AB})= S(\sigma_A) - S(\sigma_{AB})$. This implies
$S(\sigma_{AB}|| \rho^*_{AB})= S(\sigma_A) - S(\sigma_{AB})$. But from
Theorem \ref{Theorem1} we have
$S(\sigma_{AB}|| \rho^*_{AB}) - S(\sigma_A||\rho_{A}^*) \ge S(\sigma_A) - S(\sigma_{AB})$
which implies $S(\sigma_A||\rho_{A}^*)=0$ and therefore $\sigma_A =
\rho_{A}^*$ $_{\Box}$\\

It is important to note, that the lower bound derived in Lemma \ref{Lemma2} is
actually additive. This allows to draw some conclusions concerning the additivity
of the relative entropy of entanglement. In fact, for density operators that achieve
the lower bound presented in Lemma \ref{Lemma2} the relative entropy is additive.
\begin{lemma} \label{Lemma3}
If two density operators $\rho_1$ and $\rho_2$ both satisfy
$E_{RE}(\rho_i) = S(\rho_{i,A}) - S(\rho_{i,AB})$ then we have
\begin{equation}
    E_{RE}(\rho_1\otimes\rho_2) = E_{RE}(\rho_1) + E_{RE}(\rho_2) \; ,
\end{equation}
i.e. the relative entropy of entanglement is additive for $\rho_1$ and $\rho_2$.
\end{lemma}
{\bf Proof:} It is obvious that for any $\rho_1$ and $\rho_2$ we have
$E_{RE}(\rho_1\otimes\rho_2) \le E_{RE}(\rho_1) + E_{RE}(\rho_2)$. But on the
other hand $E_{RE}(\rho_1\otimes\rho_2)\ge S(\rho_{1A}\otimes\rho_{2A}) - S(\rho_{1AB}\otimes\rho_{2AB}) = E_{RE}(\rho_1) + E_{RE}(\rho_2)$ because of
additivity of the conditional entropy. Therefore
$E_{RE}(\rho_1\otimes\rho_2) = E_{RE}(\rho_1) + E_{RE}(\rho_2)$$_{\Box}$ \\

In summary we have proven a new inequality relating the quantum conditional entropy and
quantum relative entropy.
To demonstrate the usefulness of this inequality, we have used it to derive a new lower
bound on the relative entropy of entanglement as well as some remarkably simple
proofs of some other properties of the relative entropy of entanglement. Relations between
different entanglement measures could be obtained from our inequality and we
believe that this will lead to other useful applications in such
diverse fields as entanglement, distinguishability or thermodynamics.

Interesting discussions with V. Vedral are acknowledged. This work was
supported by EPSRC, The Leverhulme Trust, the EQUIP programme of the
European Union and the State Scholarships Foundation of Greece.

\end{multicols}

\end{document}